\documentclass[
    ,final            
  ]
  {aipproc}

\layoutstyle{6x9}

\newcommand{\be}{\begin{equation}}
\newcommand{\ee}{\end{equation}}
\newcommand{\bea}{\begin{eqnarray}}
\newcommand{\eea}{\end{eqnarray}}
\newcommand{\ceqn}[1]{equation~(\ref{#1})}
\newcommand{\ceq}[1]{(\ref{#1})}

\newcommand{\Inf}{{\cal I}}

\newcommand{\like}{{\cal L}}
\newcommand{\SIM}{{\em SIM}}

\begin{document}

\title{Bayesian Adaptive Exploration}

\author{Thomas J. Loredo}{
  address={Dept.\ of Astronomy, Cornell University, Ithaca, NY 14853, USA}
}

\begin{abstract}
I describe a framework for adaptive scientific exploration based on
iterating an {\em Observation--Inference--Design} cycle that allows
adjustment of hypotheses and observing protocols in response to the
results of observation on-the-fly, as data are gathered.  The framework
uses a unified Bayesian methodology for the inference and design
stages:  Bayesian inference to quantify what we have learned from the
available data and predict future data, and Bayesian decision theory to 
identify which new
observations would teach us the most.  When the goal of the experiment
is simply to make inferences, the framework identifies a
computationally efficient iterative ``maximum entropy sampling''
strategy as the optimal strategy in settings where the noise statistics
are independent of signal properties.  Results of applying the method
to two ``toy'' problems with simulated data---measuring the orbit of an
extrasolar planet, and locating a hidden one-dimensional object---show
the approach can significantly improve observational efficiency in
settings that have well-defined nonlinear models.  I conclude with
a list of open issues that must be addressed to make Bayesian
adaptive exploration a practical and reliable tool for optimizing
scientific exploration.
\end{abstract}

\maketitle


\section{Introduction}

The classical paradigm for the scientific method follows a rigid
sequence of hypothesis formation, followed by experiment and then
analysis.  It bears little resemblance to the adaptive, self-adjusting
behavior of the human brain, which learns from experience
incrementally, making decisions and adjusting questions on-the-fly.
The classical paradigm has served science well, but there are many
circumstances where what has been learned from past data could be
profitably used to alter the collection of future data to more
efficiently address the questions of interest.  In this paper I
describe an approach for developing such adaptive observing
strategies building on existing ideas from the theory of Bayesian
experimental design.  

The idea that use of partial knowledge can improve the design of
experiments has long been recognized in statistics; there are
well-developed theories of experimental design using both the
frequentist and Bayesian approaches to statistics (good entry
points to the large literature are \cite{Fedorov72, Chernoff72,AF97} for frequentist
design, and \cite{CV95, Toman99} for Bayesian design).
Unfortunately, practice has lagged theory, largely due to the
complicated calculations required for rigorous experimental design,
particularly in adaptive settings where many designs must be
calculated.  Until recently most work focused on classes
of problems that are analytically tractable (e.g., linear models with
normal errors, and, in Bayesian design, with flat or conjugate
priors).  Treatment of nonlinear models was typically handled only
approximately, by linearizing about a best-fit model.  
These limitations have
discouraged application to problems of interest to astronomers and
physicists, which often have substantial nonlinearities and other
complications.  In addition, the gains offered by optimal designs in
analytically tractable settings are often only modest.  Finally, in
these settings frequentist and Bayesian designs are the same or very
similar, suggesting (erroneously) that the two approaches have little
distinguishing themselves from each other.

In recent years computational and theoretical developments finally
enable one to undertake rigorous nonlinear Bayesian design in
complicated settings.  Most important for the approach described
here are: 
\begin{itemize}
\item The discovery that a wide and interesting class of
design problems can be analytically simplified, revealing that
the optimal observing strategy obeys a relatively
simple {\em maximum entropy sampling} criterion \cite{SW97, SW00};
\item The development of flexible and rigorous methods for Bayesian
computation based on sampling posterior distributions for models
and parameters with Markov chain Monte Carlo (MCMC) algorithms,
allowing rigorous optimal design with nonlinear models \cite{MP95,CMP95,MP96,Muller99}.
\end{itemize}
In the following section I describe the basic principles behind
Bayesian design in an adaptive setting---here dubbed {\em Bayesian
adaptive exploration} (BAE)---and highlight how recent
developments open the door to applications of realistic
complexity.  The subsequent two sections describe results of two
proof-of-concept calculations showing that BAE could improve
observational efficiency in a variety of problems in the physical
sciences and engineering.  The first example concerns measuring the orbit of an
extrasolar planet, and is motivated by the needs of the upcoming
Space Interferometry Mission (\SIM) which will astrometrically survey
nearby stars for evidence of planets, including a search for Earth-like
planets around Sun-like stars (see \cite{LC03}\ for further discussion
of the possible role of BAE for \SIM).  The second example concerns
locating a hidden one-dimensional object, and is motivated by the
need to optimally deploy a variety of sensing technologies to
efficiently and accurately locate and identify buried landmines.
The final section briefly describes the main open issues that must
be addressed to make BAE a truly useful tool for optimal
scientific exploration.

\section{Bayesian Adaptive Exploration}

BAE iterates an {\em Observation--Inference--Design} cycle depicted
in Figure~1.  In the observation stage, new data
are obtained based on an observing strategy produced by the previous
cycle of exploration.  The inference stage synthesizes the information
provided by previous and new observations to assess hypotheses of
interest.  This synthesis produces interim results such as signal
detections, parameter estimates, or object classifications.  Finally,
in the design stage the results of inference are used to predict future
data for a variety of possible observing strategies; the strategy that
offers the greatest predicted improvement in inferences (subject to any
resource constraints) is passed on to the next
Observation--Inference--Design cycle.  The observation stage
will follow the observation procedure dictated by the nature of
the experiment and need not be discussed further here.  
In the remainder of this section I outline the
components of the inference and design stages, the stages where
the tools of Bayesian statistics enter the process.

\begin{figure*}[t]
\centering
\includegraphics[width=135mm]{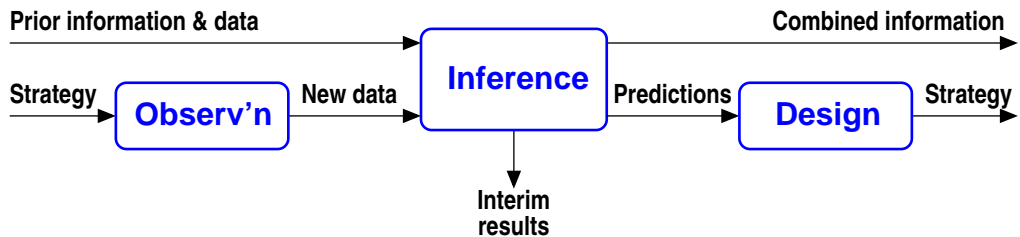}
\caption{Information flow through one cycle of the adaptive
exploration process.}
\end{figure*}

\subsection{Inference Stage}

BAE uses the tools of Bayesian inference for the inference stage.  Readers
of this volume are most likely familiar with these tools.  We review them
briefly here in order to establish notation.\footnote{For  
introductions to
Bayesian inference, see \cite{Berger85, Sivia96, Jaynes03}, and our 
{\em Bayesian Inference in the Physical Sciences} web site
(\url{http://www.astro.cornell.edu/staff/loredo/bayes/}).}

We address inference questions  by calculating probabilities for
the hypotheses of interest ($H_i$) given the available data ($D$)
and the underlying modelling assumptions ($M$); the resulting {\em
posterior probability distribution} is denoted
$p(H_i|D,M)$.  Bayes's theorem expresses it in terms of
more directly calculable probabilities as follows;
\be
p(H_i|D,M) = p(H_i|M) {p(D|H_i,M) \over p(D|M)},
\label{BT}
\ee
where $p(H_i|M)$ is the prior probability distribution for $H_i$,
$p(D|H_i,M)$ is the sampling distribution for the data, and
$p(D|M)$ is the prior predictive distribution for the data.  
In inference one is interested
in how the posterior varies with $H_i$, with the data fixed at the
observed values.  Thus what is of most interest about the sampling
distribution is how it varies with respect to the $H_i$, not the data.
Considered as a function of $H_i$, it is called the {\it likelihood}
for $H_i$, and denoted $\like(H_i) \equiv p(D|H_i,M)$.  The prior
predictive distribution has no dependence on $H_i$ and thus plays
the role of a normalization constant.  It can be calculated 
by summing the numerator in Bayes's theorem:
\be
p(D|M) = \sum_i p(H_i|M) \like(H_i).
\label{p-D}
\ee
This is just the average likelihood for $H_i$, with the prior giving
the averaging weight.  Accordingly, $p(D|M)$ is sometimes called
the average, global, or marginal likelihood.

When the hypotheses are labeled by one or more continuous parameters,
$\theta$, the probability for any hypothesis about $\theta$ can be
calculated from the posterior probability density, $p(\theta|D,E)$,
usually via an integral of some sort. Bayes's theorem holds for posterior
densities as well as for discrete distributions; the posterior density is
proportional to the product of the prior density $p(\theta|M)$ and the
likelihood function $\like(\theta)$.  The average likelihood is given by
an equation like \ceqn{p-D}, but with the sum replaced by an integral.

Several aspects of the Bayesian approach to statistical inference make it
particularly suitable for adaptive design; two are especially worth
highlighting.  First is the presence of prior probabilities. Prior
probabilities allow one to take into account information available before
embarking on an experiment. But more importantly for our purposes, they
allow for straightforward modelling of incremental, adaptive learning: the
posterior probability from current observations becomes the prior
probability for the next stage of exploration.  In this way uncertainty
can be rigorously carried through the entire exploration process,
summarizing the totality of available information in a flexible, updatable
way that can guide further exploration.

A second element of Bayesian inference particularly useful for
experimental design is the ability to {\em marginalize}, that is, to
eliminate dimensions of a problem that are not of immediate interest while
fully accounting for the effects of their uncertainty on the quantities
that are of interest.  The most common use of marginalization arises when
modelling the data requires introduction of parameters that are not
directly of scientific interest (e.g., background rates, uncertain
detector efficiencies).  Separate $\theta$ into the interesting
parameters, $\psi$, and the uninteresting {\em nuisance parameters},
$\phi$.  The inference we report for $\psi$ is the {\em marginal
distribution}, found by integrating the full posterior distribution over
the nuisance parameters,
\be
p(\psi|D,M) = \int d\phi\, p(\psi,\phi|D,M).
\label{marg}
\ee
Such calculations are useful in the inference stage because nuisance
parameters are present in nearly all real inference problems.  As we will
see below, marginalization also proves to be crucial for design because it
allows accurate prediction of future data. Having so straightforward a way
to eliminate nuisance parameters while accounting for the effects of their
uncertainty on inferences is an important advantage of the Bayesian
approach.  However, it comes with a cost: the integrals required are often
challenging, particularly when the parameter space has more than a few
dimensions.

All the probabilities above are conditional on the modelling assumptions
or background information, $M$.  It must include all information needed to
evaluate the prior and likelihood, including such things as specification
of the hypothesis space, the sample space (space of possible data sets),
and the model or models connecting these (e.g., a parameterized
theoretical model for the phenomenon of interest and distributional
assumptions about noise).  It could also include information from other
experiments.  It is important to note that $M$ must also include
specification of basic properties of the experiment (e.g., size of the
data sample, locations of samples, etc.).  In the design stage, we will be
concerned with how the posterior changes in part as a function of some of
these elements of $M$.

\subsection{Design Stage}

The goal of the design stage is to specify a new experiment (e.g., the
location in time or space for a next sample or set of samples) that will
provide new data that will best serve our scientific aims.  Of course, we
do not know what new data will actually be obtained, so our choice must
be based on predictions of what various candidate experiments might
see.  In a Bayesian framework the key ingredient for this is the {\em
predictive distribution} for the future data; Bayesian inference
provides tools for calculating this.  Once we can make predictions
about results of experiments, it only remains to choose which
experiment is best.  This is not an inferential task, but rather a
decision problem, and the proper tool for it is {\em Bayesian decision
theory} \cite{Berger85, Jaynes03}.  We now describe the prediction and
decision aspects of experimental design in turn.

A candidate experiment will be described by one or more parameters,
specifying things like sample location in space or time.  Denote these
parameters by $e$.  If we knew the parameters, $\theta$, describing the
phenomenon of interest, it would be easy to make predictions about the
data, $d$, we expect from experiment $e$; we simply evaluate the sampling
distribution for that data, $p(d|\theta,M_e)$, where the subscript on the
background information makes explicit that this information includes
specification of the sample location.  Of course, we are doing the
experiment because we do not know $\theta$.  But information about $\theta$
is available from the inference stage in the form of the posterior based
on earlier data, $p(\theta|D,M)$. A predictive distribution for $d$ that
uses this partial knowledge about $\theta$ would condition on the known
data, $D$, rather than on the unknown $\theta$; we seek $p(d|D,M_e)$. This
predictive distribution can be evaluated by introducing the unknown
$\theta$, and marginalizing;
\bea
p(d|D,M)
  &=& \int d\theta\, p(d,\theta|D,M_e)\nonumber\\
  &=& \int d\theta\, p(d|\theta,M_e) p(\theta|D,M_e),
\label{prdxn}
\eea
where for the second line we used the product rule, recognizing
that $p(d|\theta,D,M_e)=p(d|\theta,M_e)$, that is, once $\theta$
is given, conditioning on $D$ is irrelevant.  The integrand in
\ceqn{prdxn}\ is the product of two readily available quantities,
the sampling distribution for new data, and the
posterior distribution produced by the previous inference stage.

Now we must use properties of the predictive distribution to decide which
experiment---which value of $e$---is best. This requires more than just
the tools of inference.  In reaching a decision, it is not enough to
consider only the uncertainties of possible outcomes; consequences must
also be taken into account.  One might be willing to bet on an improbable
outcome if the payoff is large if it occurs and the loss is small if it
does not.  Consequences are quantified via the {\em utility} of a
decision.  The utility depends both on the possible actions we are
deciding between (to bet or not bet in the example just given) and on the
possible outcomes (determining whether the bet is lost or won).  Denoting
actions by $a$ and outcomes by $o$, the utility function defining a
decision problem can be written as a function of both, $U(o,a)$, and our
task is to decide on $a$ amidst uncertainty about $o$.

To make the best decision, decision theory specifies that one first
calculate the {\em expected utility} of the possible decisions, averaging
the utility using the probabilities of the possible outcomes:
\be
EU(a) = \sum_o p(o|I_a)\, U(a,o),
\ee
where $I_a$ denotes whatever information is available
about the outcomes, with the $a$ subscript making explicit the
possibility that the choice of action affects the probability of
the outcomes.  The best action, $\hat a$, is
the one that maximizes the expected utility,
\bea
\hat a = \arg \max_a EU(a).
\eea

In experimental design, the possible actions are the possible
experiments we might perform, indexed by $e$, and the possible
outcomes are the values of future data from that experiment, $d$.
An experimental design problem thus requires specification of
a utility, $U(d,e)$, that balances the value of $d$ for
achieving the scientific goals against possible costs of various
experiments.  Once the utility is specified, the best experiment
is the one that maximizes
\be
EU(e) = \int dd\, p(d|I_e)\, U(d,e).
\label{EU-basic}
\ee
The conditioning information most typically includes a parameterized
model for the phenomenon producing the data, and possibly values
of already collected data; so the probability we need is the
predictive distribution, $p(d|D,M_e)$.
Using \ceqn{prdxn}, we can write the expected utility as,
\be
EU(e) = \int d\theta\, p(\theta|D,M_e) \int dd\, p(d|\theta,M_e) U(d,e).
\label{EU}
\ee

To proceed we must specify $U(d,e)$.  In disciplines such as econometrics
or biometrics, there are obvious costs and benefits of decisions;
consequently, decision theory is more prominent in these disciplines than
in the physical sciences, where the goal of a study is usually not to
reach a formal decision, but to report the implications of the data for
various hypotheses.  That is, in the physical sciences we are primarily
concerned with inference, with reporting the information conveyed by the
entire posterior distribution, not with achieving a specific goal
with an assigned value.  But we can still use decision theory to
determine the best experiment to perform if we can come up with a
utility that measures how well the experiment improves our inferences.

In 1956, Lindley described how one could use tools from information
theory in a Bayesian framework to compare experimental designs when
one's purpose is simply to gain knowledge about a phenomenon
\cite{Lindley56}.  He later incorporated these ideas into the more
general theory of Bayesian experimental design outlined above, first
described in his influential 1972 review of Bayesian statistics \cite{Lindley72}.
Although non-Bayesian methods for optimal design predate Lindley's work
(standard references are \cite{Fedorov72, Chernoff72,AF97}), the
Bayesian approach provides a more fundamental rationale for many
earlier methods, and unifies and generalizes them (see \cite{CV95}\ for
discussion of the relationships between Bayesian and non-Bayesian
design).

Lindley suggested that if the goal is to learn about $\theta$, a natural
utility function is the information in the final posterior distribution
for $\theta$, as measured by information theory,
\be
U(d,e) = \int d\theta\, p(\theta|d,D,M_e) \log p(\theta|d,D,M_e),
\label{U-Shannon}
\ee
where $p(\theta|d,D,M_e)$ is the posterior for $\theta$ {\em including
future data} $d$, and the right hand side is just the negative Shannon
entropy of this distribution.\footnote{For a Gaussian distribution with
standard deviation $\sigma$, the
Shannon entropy is
proportional to $-\log(\sigma)$ and thus increases with decreasing
$\sigma$ as one would expect of a measure of information; but it is a 
more general measure of spread
than the standard deviation.  To be formally correct, the argument of
the logarithm in \ceqn{U-Shannon}\ should be divided by a measure on the
parameter space so the argument is dimensionless; this has no
significant effect on our results.  An alternative definition of
information is the cross-entropy or Kullback-Leibler divergence between
the posterior and prior; it gives the same results as the Shannon
entropy for this calculation \cite{MacKay92}.}  
Using \ceqn{EU}, the expected information from experiment $e$ is,
\bea
EU(e) 
 &=& \int d\theta\, p(\theta|D,M_e) \int dd\, p(d|\theta,M_e) \nonumber\\
 &&\times \int d\theta'\, p(\theta'|d,D,M_e) \log p(\theta'|d,D,M_e).
\label{EU-Shannon}
\eea
The best experiment is the one that maximizes the expected information.
If there are definite, variable costs associated with various choices
of $e$, they can be subtracted from the utility.  Up to this possible
generalization, we have completed specification of the design stage.

\subsection{Implementation}

In the three decades since Lindley advocated designing to maximize
information, the theory of design has matured significantly.  But as noted
in Toman's recent review of Bayesian design, ``unfortunately much of the
work in this area remains purely theoretical'' \cite{Toman99}.  This is
largely due to the computational complexity of Bayesian design, an
obstacle noted already in Lindley's foundational work.  In experimental
design, one must account for both uncertainty regarding the hypotheses
under consideration, and uncertainty about the values of future
data---\ceqn{EU}\ has integrals over both $\theta$ and $d$.  For the
former, one must perform the difficult parameter space integrals that are
characteristic of Bayesian inference \cite{Loredo99}; for the latter, one
must additionally integrate over the sample space as is typically done in
frequentist calculations (e.g., by Monte Carlo simulation of data).  In a
sense, experimental design is the arena in which the Bayesian and
frequentist outlooks meet, producing problems with the combined complexity
of both approaches.  In addition, when designing to maximize information,
the utility itself requires a nontrivial parameter space integration (the
integral over $\theta'$ in \ceqn{EU-Shannon}), adding to the complexity.

The computational complexity of Bayesian experimental design has led
researchers to focus on problems where all or most of the needed
integrals are analytic, e.g., data with additive Gaussian noise and
linear models.  In particular, little work exists rigorously handling
nonlinear models (most nonlinear design work relies on linearization
about a best-fit model; e.g., \cite{MacKay92, SS98}).  For linear
models with additive Gaussian noise, it is known that design does not
depend on the values of the available data, only on their noise levels
and noise correlations \cite{MacKay92} (this is because the data values
affect the location but not the width of the posterior for linear
models, and Shannon information measures the width). This limits the
improvements one can achieve using optimal designs with such models.

In recent years, two advances have set the stage for rigorous Bayesian
design with nonlinear models.  On the theoretical front, investigators
have found a broad and interesting class of problems for which the
expected information expression in \ceqn{EU-Shannon}\ can be analytically
simplified, reducing the dimensionality of the integrals needed.  On
the computational front, investigators have used the technique of
posterior sampling, particularly via MCMC
methods, to calculate and report quantities needed for nonlinear
design without any approximation of the integrands.  We review
these developments in turn.

\subsubsection{Maximum Entropy Sampling}

Let us explore the structure of the expected information expression
analytically.  Information is a functional, a mapping from functions to
real numbers.  Our manipulations will involve the information in various
distributions, so we need a notation that is both compact but clearly
shows what distribution is being used for an information calculation.
Let $\Inf[x|I]$ denote the information in the distribution $p(x|I)$, so that
\be
\Inf[x|I] \equiv \int dx\, p(x|I) \log p(x|I).
\label{inf-def}
\ee
With this notation, the expected utility, \ceqn{EU-basic}, can
be written,
\be
E\Inf(e) = \int dd\, p(d|I_e) \Inf[\theta|d,I_e].
\label{EI-e}
\ee

The integrand of \ceqn{EI-e}\ includes the information in the
posterior distribution.  This can be rewritten in terms of the
information in other distributions using Shannon's theorem, the
information theory analog to Bayes's theorem.  As with the proof of Bayes's
theorem, we establish Shannon's theorem by looking at the joint
distribution for $\theta$ and $d$, and using the product rule to factor
it.  Dropping the common conditioning proposition $I_e$ for the moment,
\bea
\Inf[d,\theta]
  &=& \int dd \int d\theta\, p(d,\theta) \log p(d,\theta)\nonumber\\
  &=& \int dd \int d\theta\, p(d,\theta) \log p(\theta)
      + \int dd \int d\theta\, p(d,\theta) \log p(d|\theta)\nonumber\\
  &=& \Inf[\theta] + \int d\theta\, p(\theta)\Inf[d|\theta],
\label{ST-1}
\eea
where for the first term we used the fact that $d$ appears only in a
$p(d|\theta)$ factor in the first factor of the second line; this
integrates to unity.  Repeating the calculation but switching the order of
factoring $d$ and $\theta$ we similarly find,
\be
\Inf[d,\theta] = \Inf[d] + \int dd\, p(d)\Inf[\theta|d].
\label{ST-2}
\ee
Equating the right hand sides of equations \ceq{ST-1}\ and \ceq{ST-2}\ gives
Shannon's theorem.  But note that the
last term in \ceqn{ST-2}\ is just the expected
information, \ceqn{EI-e}.  Thus Shannon's theorem shows that
\be
E\Inf(e) = \Inf[\theta|I_e] + \int d\theta\, p(\theta|I_e)\Inf[d|\theta,I_e]
   - \Inf[d|I_e].
\label{ST}
\ee
In words, the expected information from an experiment is the
information in the prior, plus the average information in the
sampling distribution, minus the information in the predictive
distribution.

In most cases the information in the prior (i.e., the posterior from
consideration of existing data) will not depend on the sampling scheme for
future data, so the first term in \ceqn{ST}\ will be constant. More
subtly, in many cases the information in the sampling distribution will
also be independent of the sample location. For example, often the
sampling distribution describes the influence of noise on the data, with
noise properties that are independent of the level of whatever signal the
noise is added to (this would be true of many experiments with additive
noise, but not of experiments with Poisson statistics except in the limit
where background dominates signal).  In these cases, the second term is
also constant.  As a result, only the last term depends on $e$; writing it
out, we have,
\be
E\Inf(e) = C - \int dd p(d|D,M_e) \log p(d|D,M_e).
\label{EI-ent}
\ee
The best experiment maximizes this expression.  Noting the
minus sign, this means that the best experiment is the one
whose predictive distribution has maximum entropy.  Sampling
according to this criterion is thus called {\em maximum entropy
sampling}; despite its simplicity this criterion was discovered
in its full generality only recently by Sebastiani and Wynn \cite{SW97, SW00}.
Entropy is large for distributions that are broad and uninformative.
Thus this is an eminently reasonable criterion; it tells us that 
{\em we will learn the most by sampling where
we know the least}.

Besides providing an intuitive understanding of what Bayesian design
accomplishes, \ceqn{EI-ent}\ significantly simplifies the needed
computations.  The predictive distribution appearing in \ceqn{EI-ent}\ requires 
a parameter space integral, so calculating the $e$-dependent part
of $E\Inf(e)$ requires nested sample space and parameter space integrals. 
But the full expected information expression, \ceqn{EU-Shannon}, involves
a further nested parameter space integral.  When the conditions for
maximum entropy sampling apply, we are spared of one level of parameter
space integration.

\subsubsection{Posterior Sampling}

Even when the design problem simplifies to a maximum entropy sampling
problem, we must evaluate \ceqn{EI-ent}\ as a function of $e$.  In
situations such as those described below where BAE is being implemented
sample-by-sample, the $d$ integration will be of low dimension (perhaps
just a single dimension, as in the examples below).  But the predictive
distribution appearing in \ceqn{EI-ent}\ requires a parameter space integration
that can be challenging to do with models that have more than a few dimensions.
Since about 1990, enormous progress has been made in finding methods to
perform such integrals numerically without having to approximate the 
integrands.  The most influential methods use {\em posterior sampling}.
One creates a psuedorandom number generator that produces parameter
values sampled from the full posterior distribution.  Once a set of
such values is available, one can estimate many quantities of interest
by simple manipulation of the samples.  The samples themselves also
provide a visually appealing display of the structure of the posterior.
The most popular and most flexible class of methods for obtaining
posterior samples are MCMC methods, though for problems of small
or moderate dimension simpler methods can be feasible.

As an example of how posterior sampling can facilitate nonlinear design,
imagine we have an algorithm that provides us with a set of $N$ samples,
$\{\theta_i\}$, from the posterior for $\theta$.  Then the value
of the predictive probability density for a particular value of $d$ can be
estimated as follows;
\bea
p(d|D,M_e)
  &=& \int d\theta\, p(\theta|D,M_e) p(d|\theta,M_e)\nonumber\\
  &\approx& {1\over N} \sum_i p(d|\theta_i,M_e),
\label{prdxn-MC}
\eea
that is, simply average the values of the sampling distribution for $d$
conditioned on the posterior samples of $\theta$.  Call this estimate
$\tilde P(d)$.  In addition to calculating the {\em value} of the
predictive density, the posterior samples allow us to {\em sample} a $d$
value from it.  We simply iterate the following two steps to generate,
say, $M$ samples, $\{d_j\}$:
\begin{itemize}
\item Sample the posterior distribution, yielding a parameter point, 
$\theta_i$.  (If a pool of posterior samples is already available, just
sample uniformly from that pool.)
\item Sample $d_j$ from the sampling distribution for $d$ conditioned
on $\theta_i$, $p(d|\theta_i,M_e)$.
\end{itemize}
The last step is often trivial; e.g., for data modeled as a signal
with additive noise with a normal distribution, one simply draws a
sample from the normal distribution (suitably scaled and shifted).

With these two ingredients, we can estimate \ceqn{EI-ent}\ in the
familiar Monte Carlo way;
\be
E\Inf(e) \approx {1\over M} \sum_j \log \tilde P(d_j),
\label{EI-MC}
\ee
where we have dropped the uninteresting constant.  This is the
approach used for the calculations described below, evaluating
$E\Inf(e)$ over a grid of $e$ values and locating the maximum
directly.  As noted in the final section, other investigators
have developed other algorithms that inventively use posterior
sampling, in some cases to ``sample'' over the $e$ dimension of
the problem in a way that targets attention to the optimal design.

The combination of analytical simplification via maximum entropy
sampling (where possible) and use of posterior sampling algorithms
is finally making rigorous nonlinear design feasible.  We now
turn to some simplified but nontrivial examples to explore the potential
of BAE to improve the effectiveness of exploration in an adaptive framework.

\section{Example: Extrasolar Planet Measurements}

Consider the problem of optimally scheduling
observations of a star in order to characterize the orbit of a planet
detected via radial velocity measurements of the Keplerian reflex motion
of the star.  The data are modeled by
\begin{equation}
d_i = V(t_i; \tau, e, K) + n_i,
\label{d=Ve}
\end{equation}
where $n_i$ is the noise contribution to datum $i$, and $V(t_i;
\tau, e, K)$ gives the Keplerian velocity along the line of site
as a function of time $t_i$ and of the orbital parameters $\tau$
(period), $e$ (eccentricity), and $K$ (velocity amplitude); for
simplicity three purely geometric parameters are suppressed. This
function is strongly nonlinear in all variables except $K$.  When
the eccentricity vanishes, it is simply a sinusoid; for nonzero
eccentricity it has a more complicated periodic shape.  Our
goal is to learn about the parameters $\tau$, $e$ and $K$.

Figure 2 shows results from one Observation--Inference--Design cycle,
using simulated data. 
Figure~2a depicts the initial observation stage.  The points show the data 
from a ``setup'' observation;
observations were made at 10 equispaced times, and the error bars
indicate the noise standard deviation (the noise distribution is Gaussian
with zero mean and $\sigma=8$~m~s$^{-1}$).  The curve shows
the true orbit with typical exoplanet parameters ($\tau=800$~d,
$e=0.5$, $K=50$~ms$^{-1}$).  

Figure~2b shows some
results from the inference stage using these data.  Shown are 100
samples from the marginal posterior density for $\tau$ and $e$.
In a more careful calculation, we would use more samples and
smoothing to find contours of credible regions; here it suffices to
note that the displayed cloud of points should conservatively bound a
90\% credible region.  The period and eccentricity are
usefully constrained by the 10 data points, but there is significant
uncertainty that would not be well described
by a Gaussian distribution (even correlated).   Figure~2c shows how easily a
complicated marginal distribution can be found using the samples; it
displays the marginal distribution for two interesting physical parameters
of the system, the planet's semimajor axis,
$a$, and $m\sin i$, the product of its mass and the sine of its orbital
inclination.  These are each nonlinear functions of the three model
parameters.  To produce Figure 2c we simply evaluated these functions
for each of the 100 samples of $(\tau, e, K)$ already produced; this is
much simpler than numerically evaluating the multiple integral defining
the marginal distribution over a $(m\sin i,a)$ grid.  By reporting the
actual sample values, other investigators could use the results of
these observations in their own calculations and fully account for the
uncertainties simply by evaluating any quantities of interest over the
set of samples.

Figure~2d illustrates the design
stage.  The thin curves display the uncertainty in the predictive
distribution as a function of sample time; they show
the $V(t)$ curve associated with 15 of the parameter
samples from the inference stage.  The spread of these curves at
a particular time displays the uncertainty in the predictive distribution
at that time.  The Monte Carlo calculation
of the actual expected information using all 100 samples is plotted as the thick curve
(right axis, in bits, offset so the minimum is at 0 bits).  The curve
peaks at $t=1925$~d, the time used for observing in the next cycle.

\begin{figure}[t]
\begin{tabular}{c}
\includegraphics[angle=270,scale=0.34]{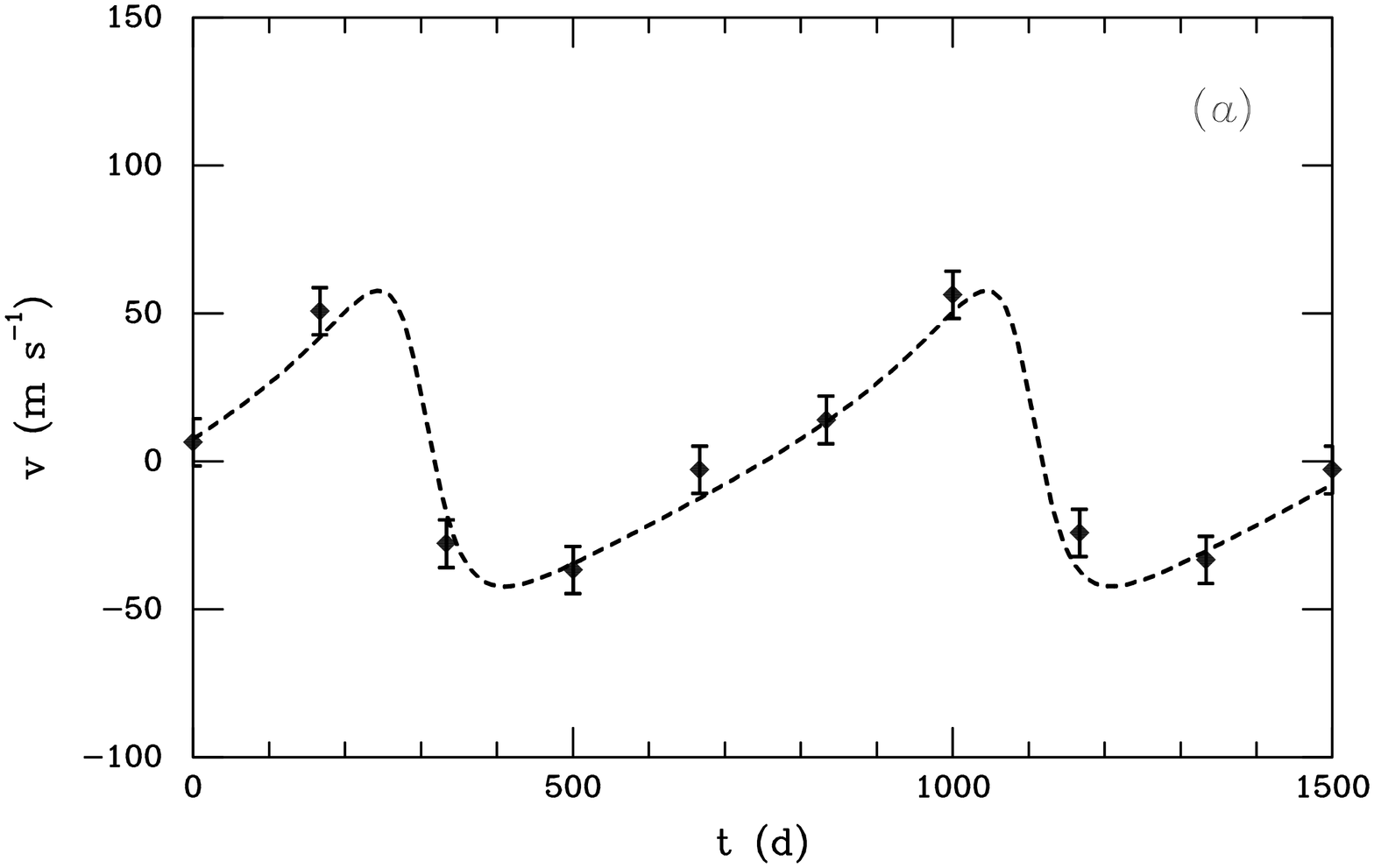}\\
\includegraphics[angle=-90,scale=0.57]{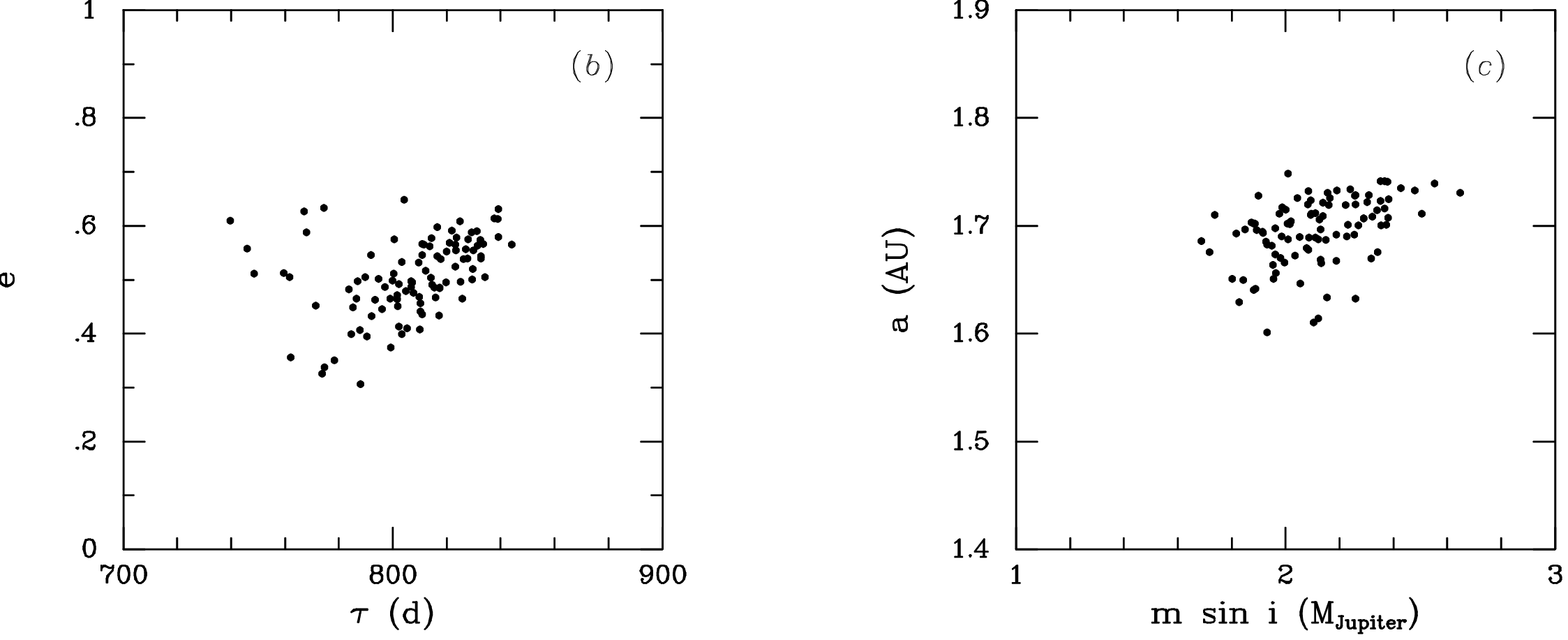}\\
\includegraphics[angle=-90,scale=0.34]{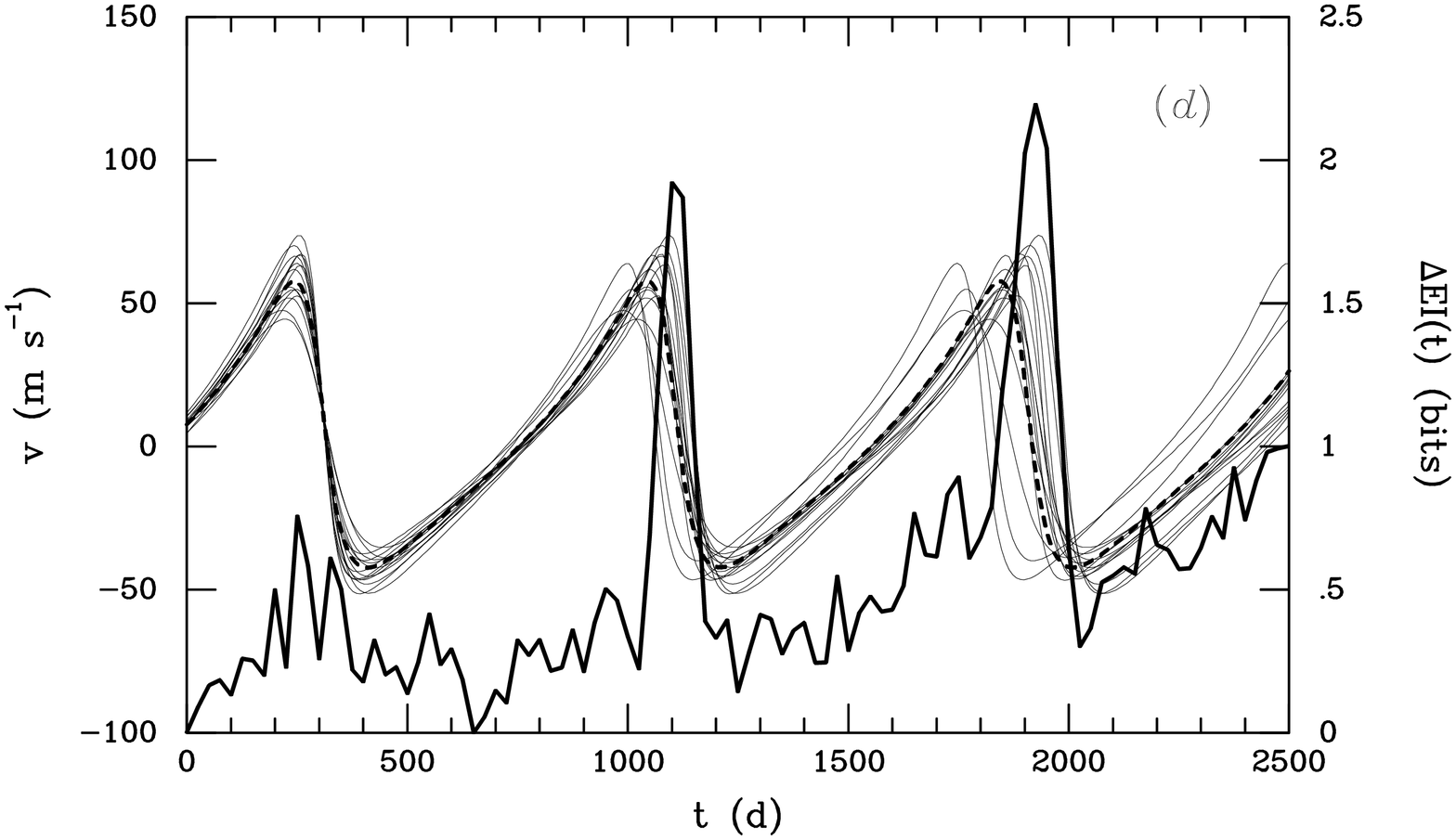}
\end{tabular}
\caption{One cycle of the exploration process characterizing the orbit of an extrasolar planet with simulated radial velocity
observations.  (a) Observation stage, showing 10
simulated observations and true velocity curve (dashed).  (b,c)
Inference stage, showing samples from the posterior distribution for
two velocity curve parameters (b) and two derived orbital parameters
(c).  (d) Design stage, showing predicted velocity curves (thin solid
curves), true velocity curve (dashed curve), and the expected
information gain for a sample at each time (thick solid curve, right
axis).}
\end{figure}

\begin{figure*}[t]
\centering
\includegraphics[angle=-90,scale=0.35]{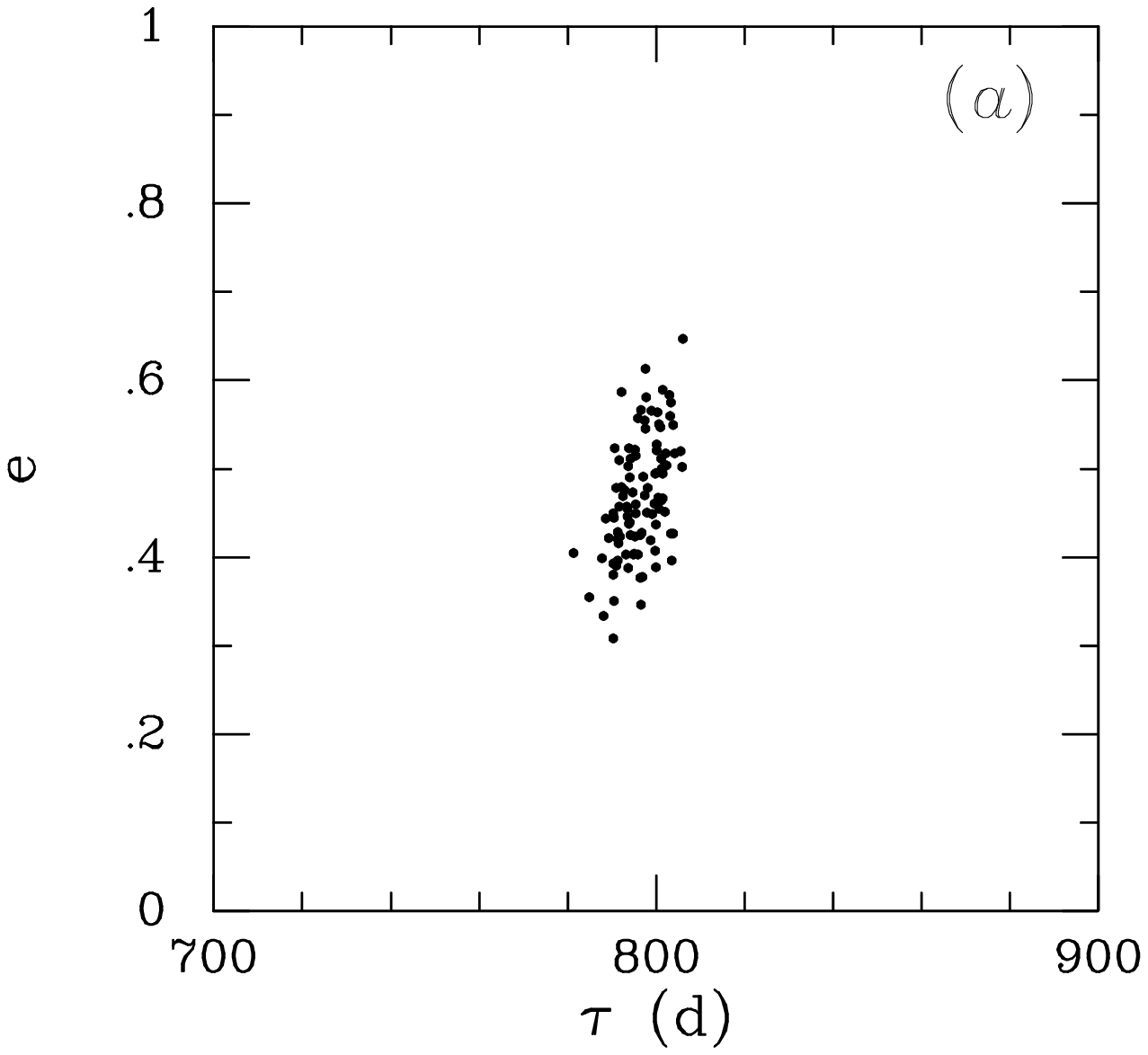}\qquad
\includegraphics[angle=-90,scale=0.35]{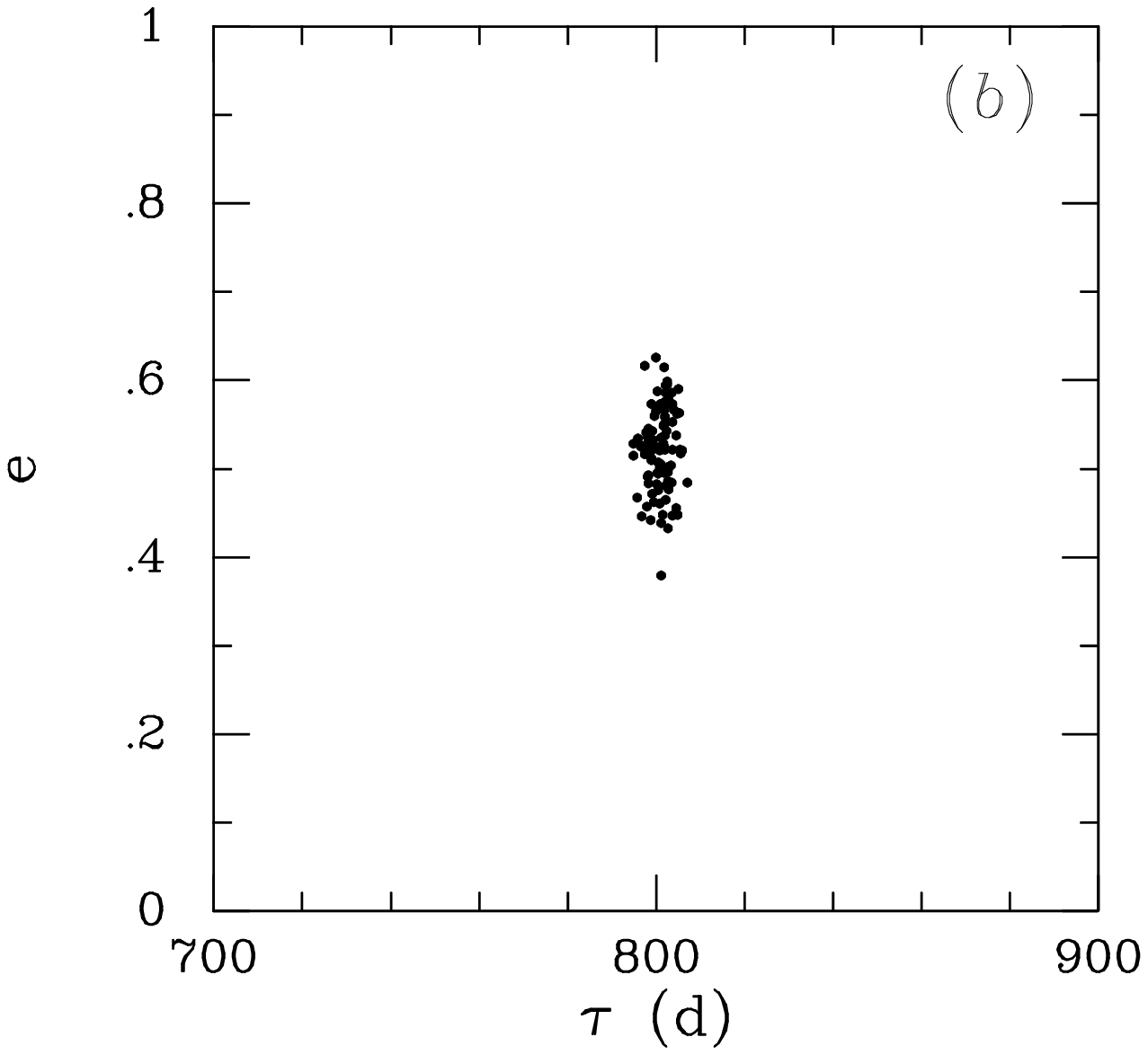}\qquad
\includegraphics[angle=-90,scale=0.35]{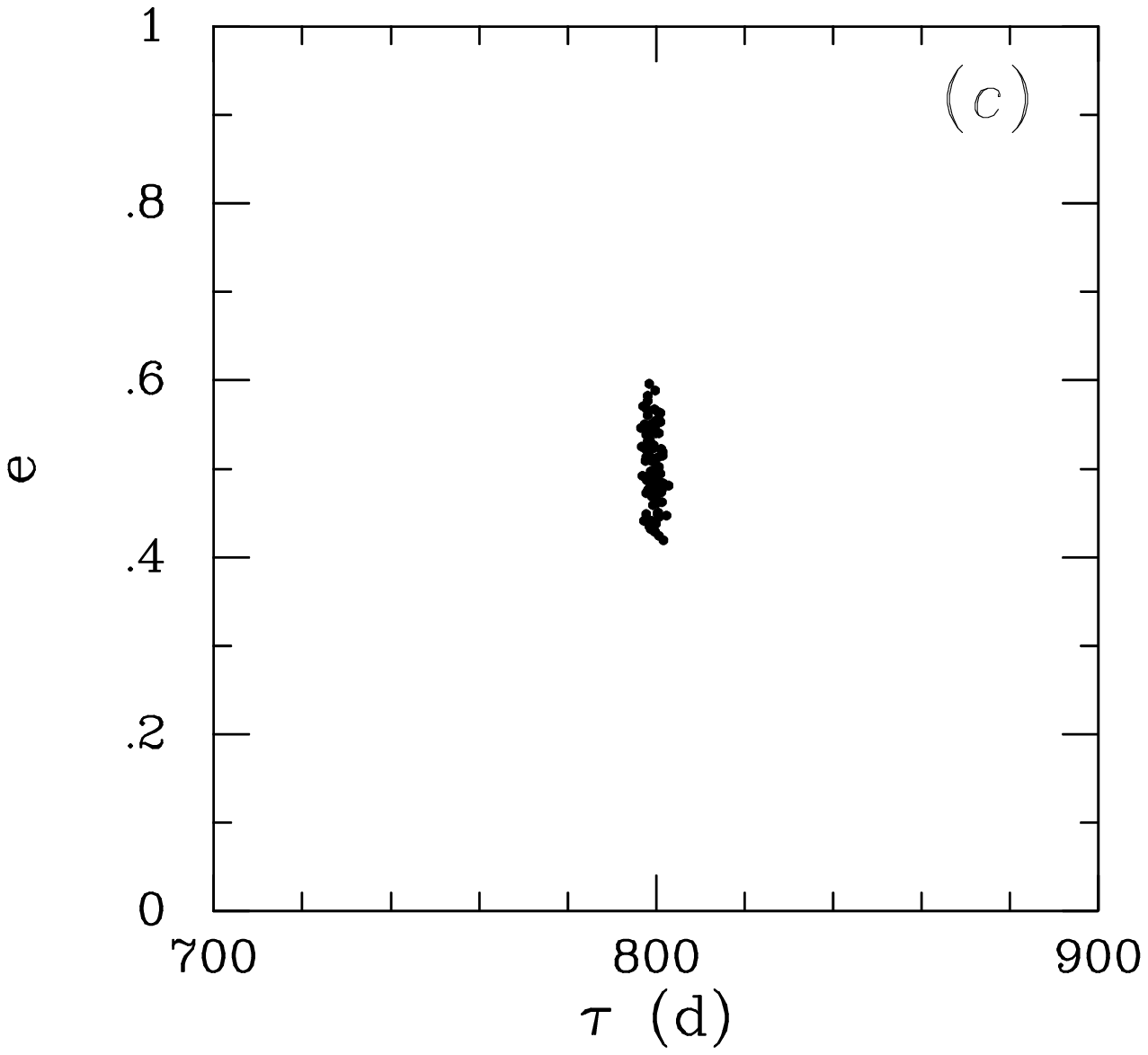}
\caption{Inference stage results from three observation-inference-design cycles
subsequent to that in Fig.~2, displaying rapid improvement of inferences.}
\end{figure*}

Figure~3a shows interim results from the inference stage of the next cycle
after making a single simulated observation at the optimal time. The period
uncertainty has decreased by more than a factor of two, and the product of
the posterior standard deviations of all three parameters (the ``posterior
volume'') has decreased by a factor
$\approx 5.8$; this was accomplished by incorporating the information {\em
from a single well-chosen datum}.  Figures~3b and 3c show similar results from
the next two cycles.  The posterior volume
continues to decrease much more rapidly than one would expect from
the random-sampling ``$\sqrt{N}$ rule'' (by factors of $\approx 3.9$ and $1.8$).
Also, the correlation in the posterior is greatly reduced and is
negligible in the final cycle.

\section{Example: Finding a Hidden Object}

Figure~4 provides a further example motivated by the problem of
detecting buried landmines using a mix of technologies---inexpensive
but noisy ferromagnetic scans, and more costly but more sensitive
acoustic scans using laser doppler vibrometry (see \cite{XG03}\ for a
discussion of some of this technology in a Bayesian setting).  Figure~4a shows a
hidden 1-d Gaussian-shaped ``object'' (dashed curve; peak at $x_0=5.2$, amplitude $A=7$,
$\hbox{FWHM}=0.6$) barely detected in an initial scan with 11 crude
($\sigma=1$) observations spaced well over a full-width apart.
Figure~4b shows samples from the marginal posterior density for $A$ and
$x_0$ from the first inference stage, displaying very substantial
uncertainty.  BAE proceeds, designing for subsequent more sensitive
observations with reduced noise level ($\sigma=1/3$).  
The design stage produces the entropy curve shown in Figure~4c (bold
curve, right axis), and specifies observing near
the best guess for the peak.

Figure~5 shows results from the inference and design stages of three subsequent
BAE cycles.  Incorporating data from a single observation taken as
specified by Cycle~1 produces the
more concentrated but complicated Cycle~2 inference of Figure~5a.  Its
``U'' shape reflects the fact that the observations constrain the amplitude
of the Gaussian at a particular point, but not the actual location of
the Gaussian.  Figure~5b shows the Cycle~2 design stage results,
specifying an observation to the left of the estimated peak.  Observing
here produces the Cycle~3 inference in Figure~5c.  Now the design
stage directs attention to the other side of the object, as indicated
by the entropy curve in Figure~5d.  The subsequent Cycle~4 inference 
is shown in Figure~5e and is impressively accurate and uncorrelated.
The posterior volume decreases by factors of $\approx 8.2$, 6.6, and
5.6 between cycles, far more dramatically than expected from random
sampling (even adjusting for the fact that only two of the original
samples lie in the signal region).
If for the last step one samples just
a few tenths of a unit from the optimal point, the nonoptimal Cycle~4
inference in Figure~5f results; strong correlations remain, and the posterior volume is
40\% larger than what was obtained using an optimal observation.

\begin{figure*}[t]
\begin{tabular}{cc}
\multicolumn{2}{c}{\includegraphics[angle=270,scale=0.35]{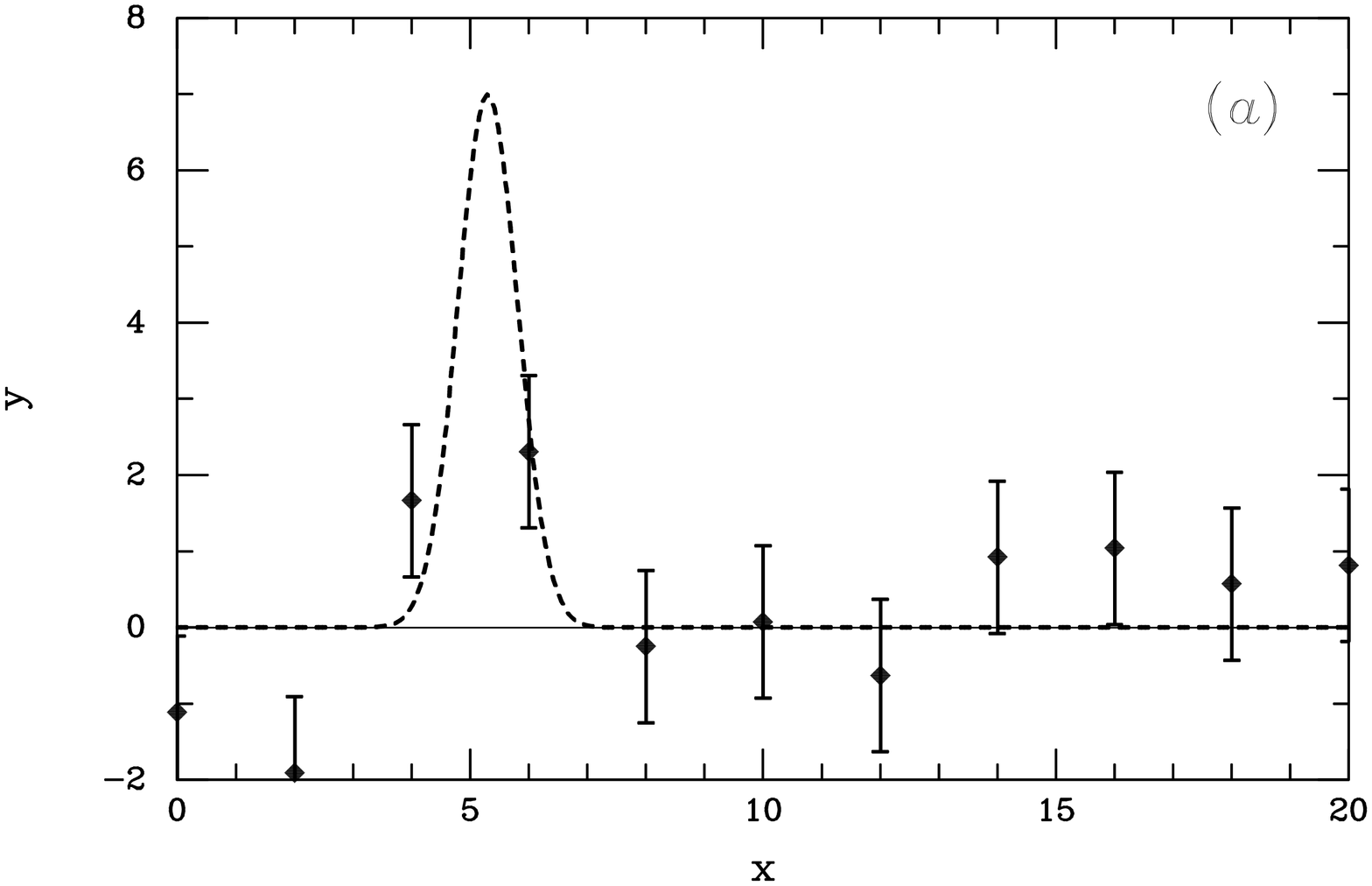}}\\
\includegraphics[angle=-90,scale=0.58]{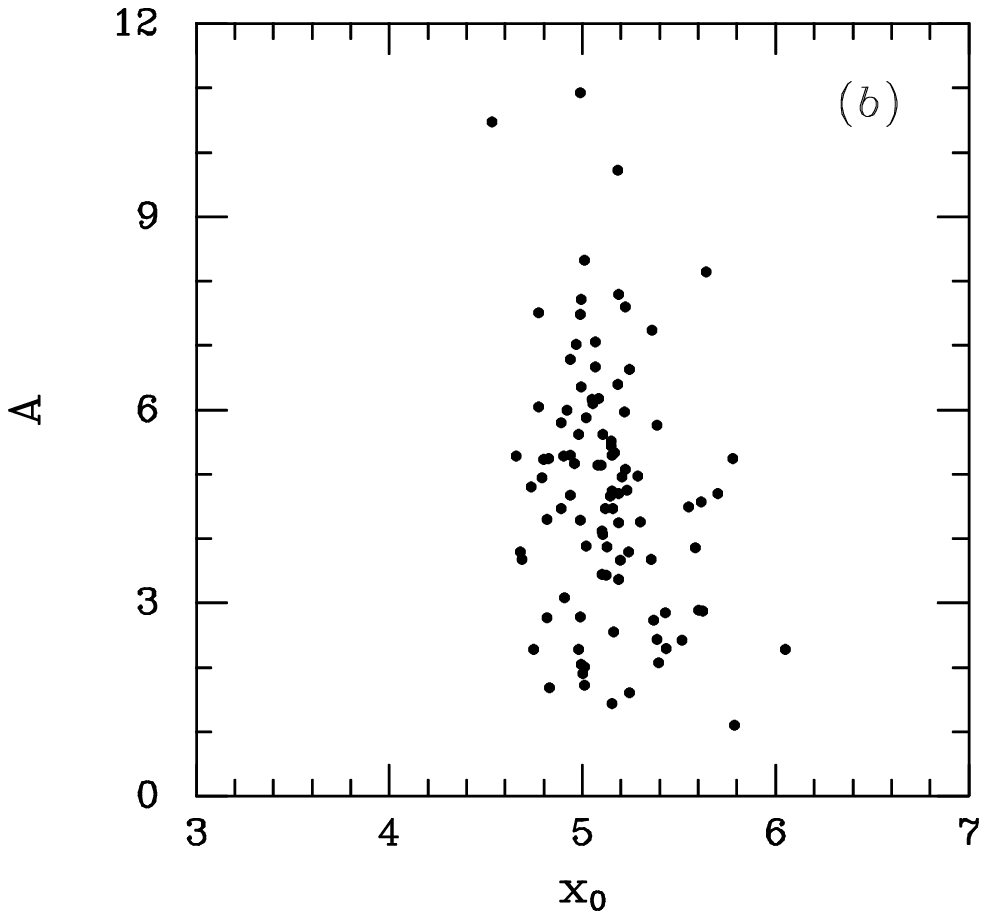}
&\includegraphics[angle=-90,scale=0.35]{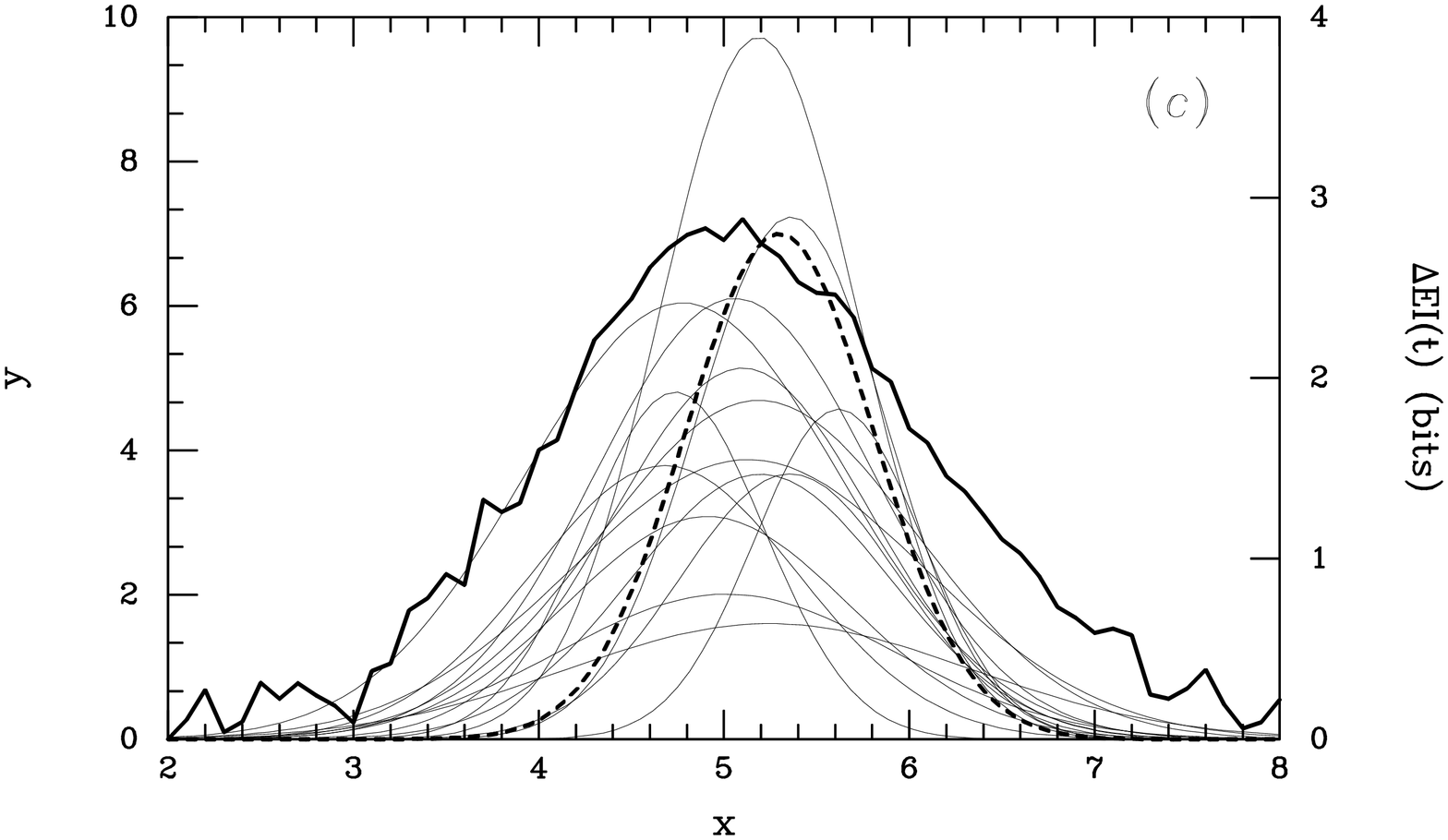}
\end{tabular}
\caption{Results from the first observation-inference-design cycle for a
simulated experiment
characterizing a hidden 1-d Gaussian object with noisy observations.}
\end{figure*}

\begin{figure*}[t]
\begin{tabular}{cc}
\includegraphics[angle=-90,scale=0.58]{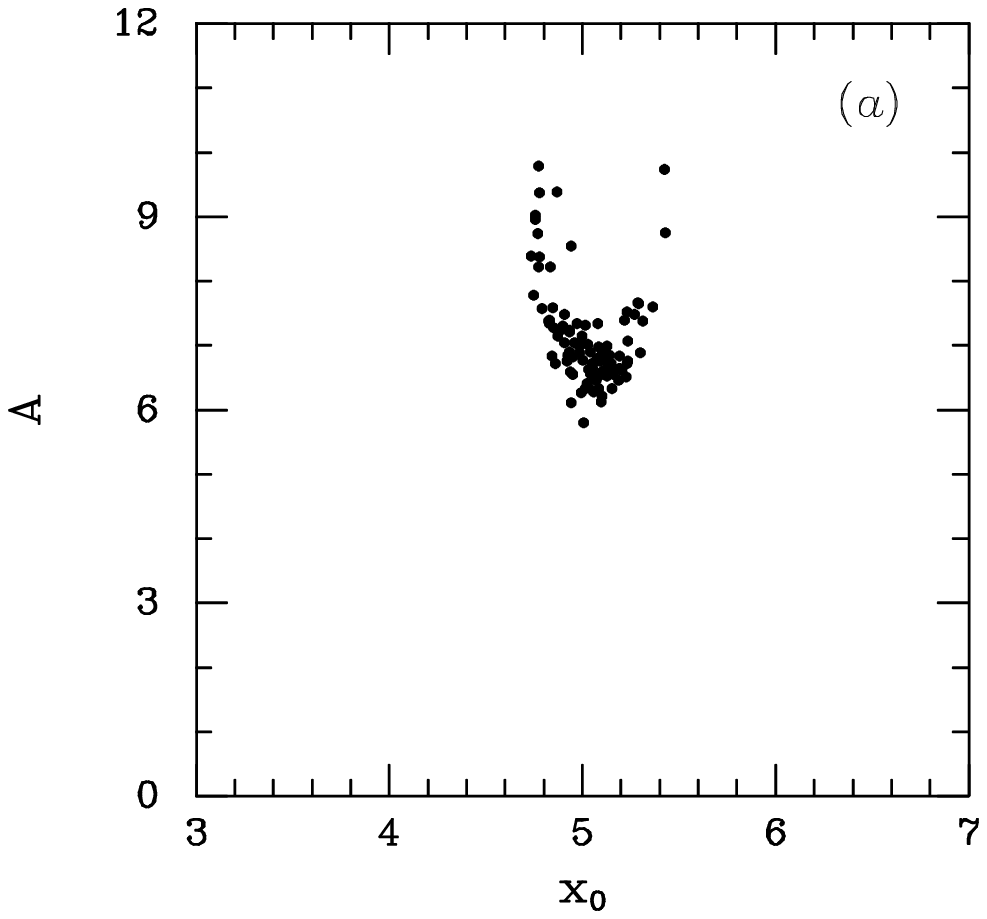}
&\includegraphics[angle=-90,scale=0.35]{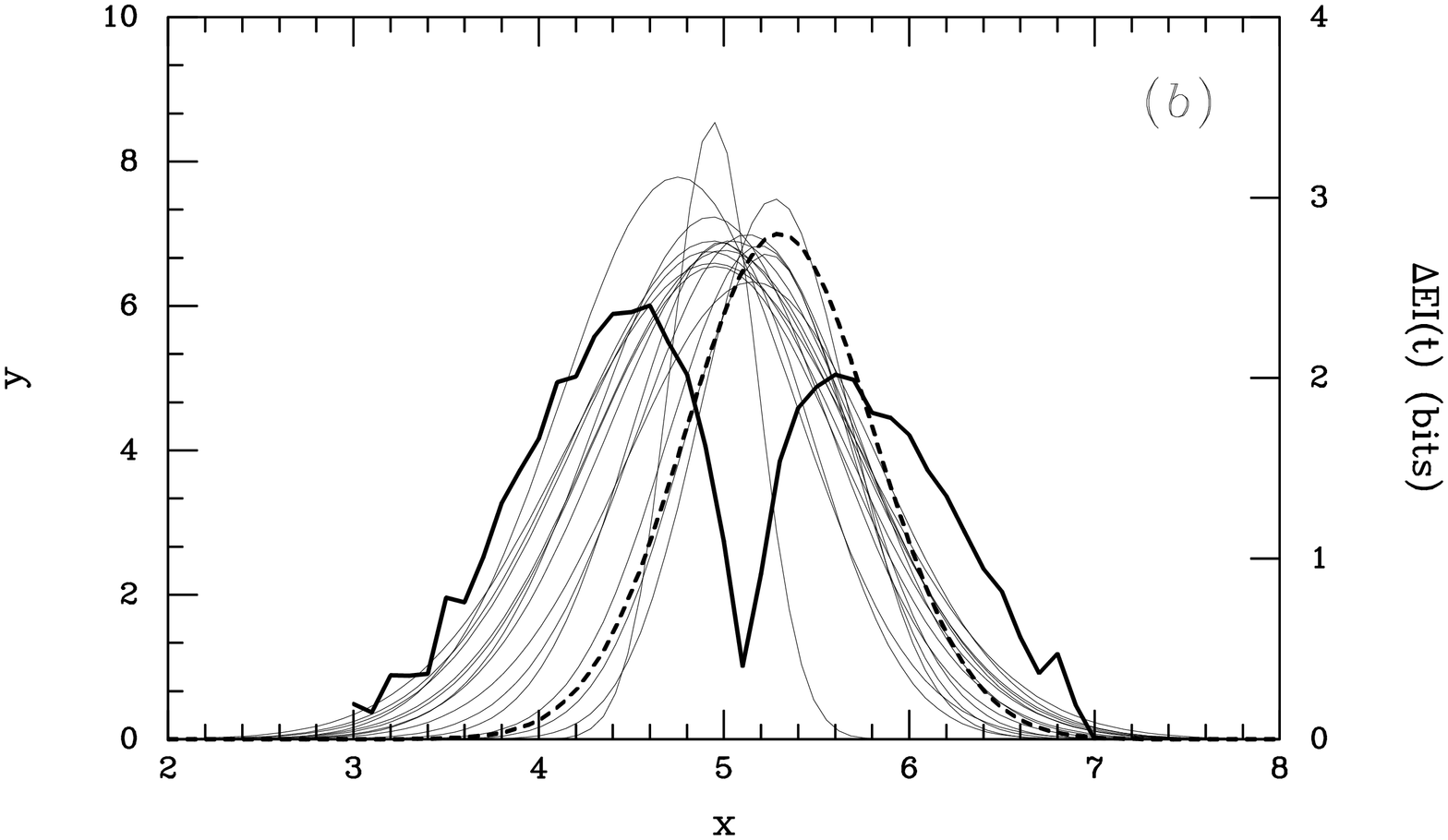}\\
\includegraphics[angle=-90,scale=0.58]{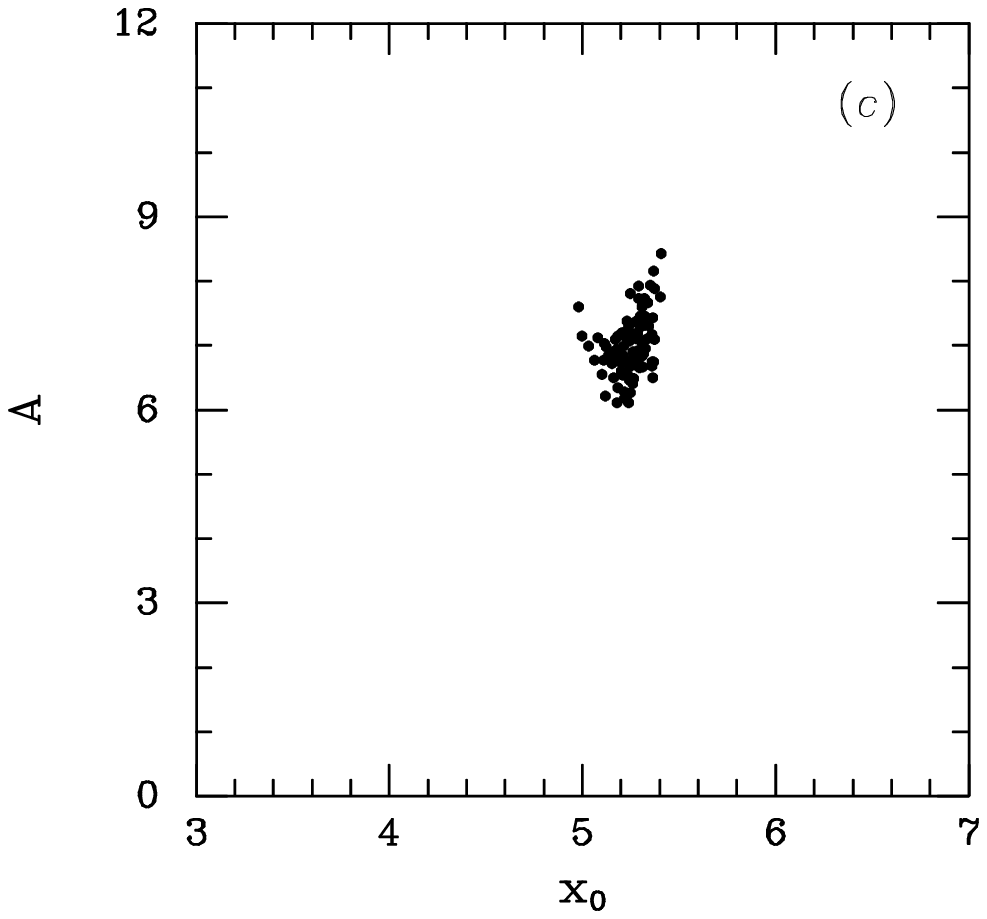}
&\includegraphics[angle=-90,scale=0.35]{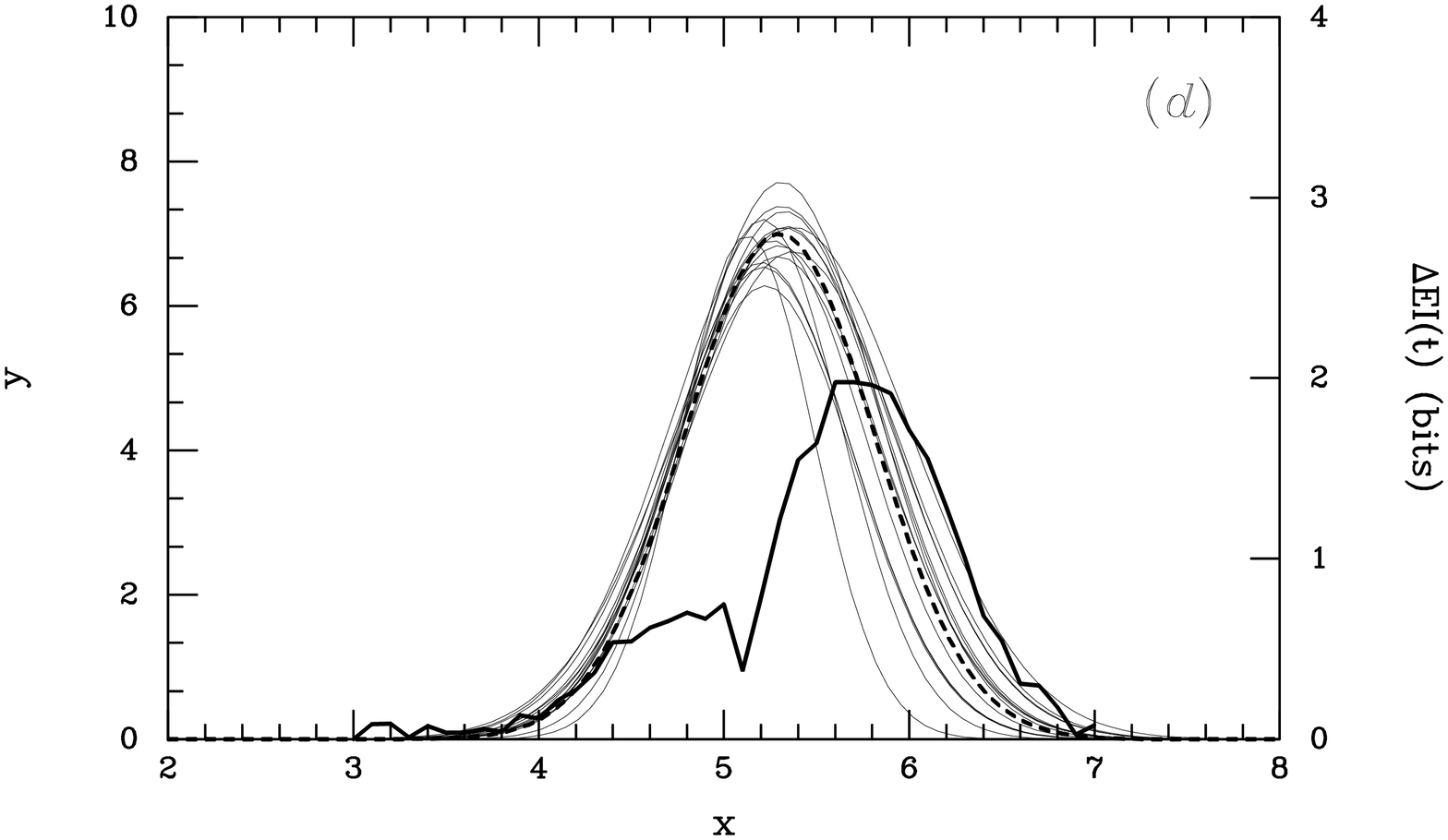}\\
\includegraphics[angle=-90,scale=0.58]{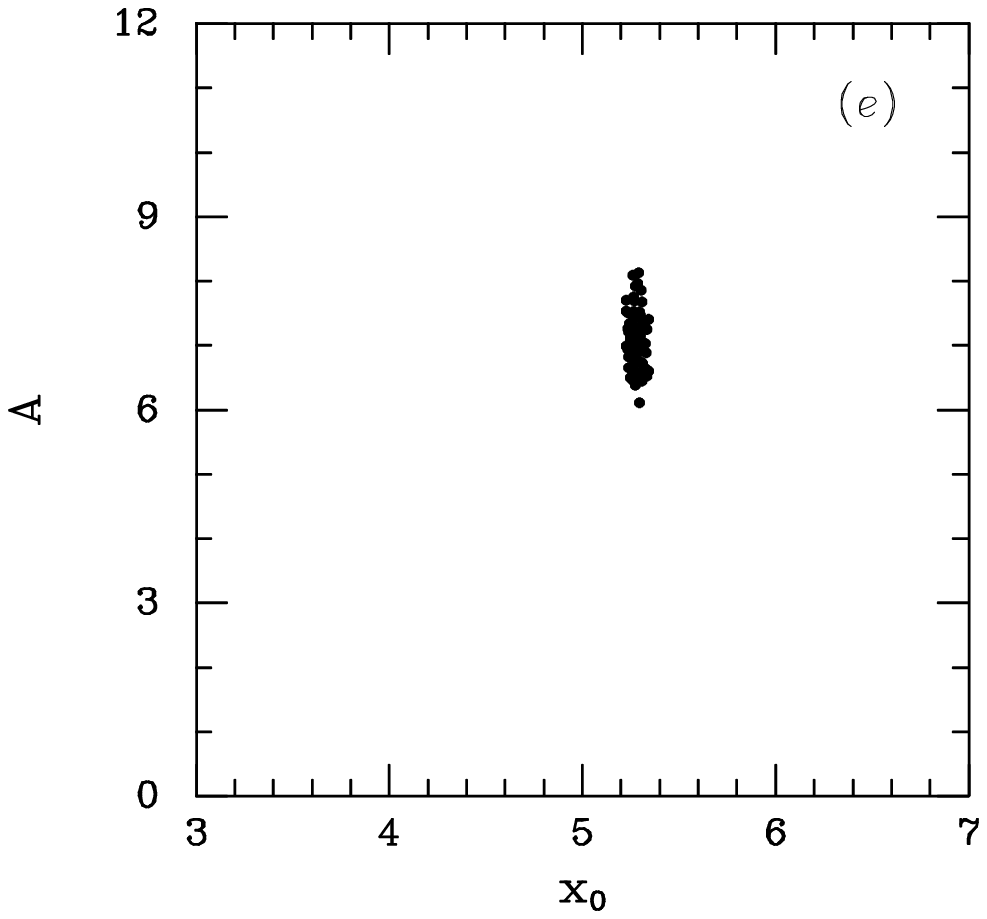}
&\includegraphics[angle=-90,scale=0.58]{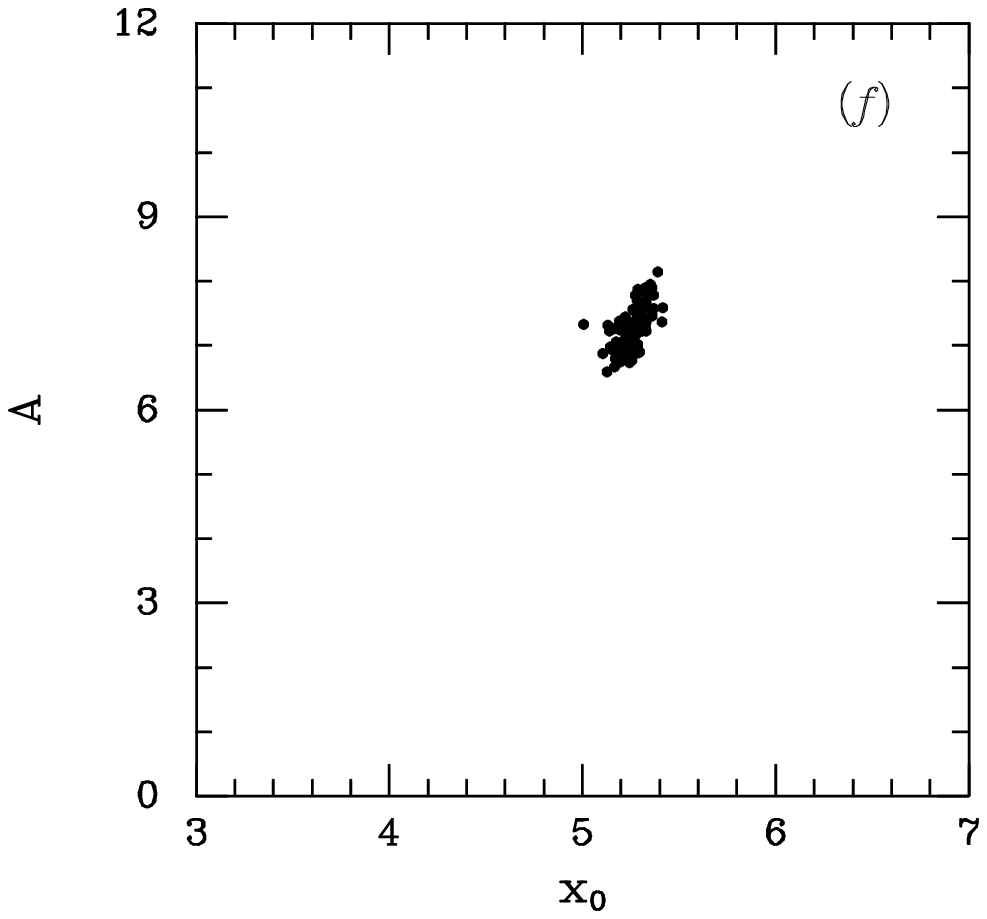}
\end{tabular}
\caption{Results from the inference and design stages of three 
observation-inference-design cycles subsequent to the cycle shown
in Fig.~4.  Panel (f) shows inference stage results in the final
cycle if a nonoptimal observation is used instead of the one
specified in the previous design stage.}
\end{figure*}

\section{Open Issues}

The examples illustrate the BAE methodology and demonstrate its potential
to very significantly improve observational efficiency in problems where one
can adjust the sampling strategy on-the-fly to make inferences about 
well-specified nonlinear models. But several issues need to be addressed
to make BAE useful in realistically complicated settings.  The field of
experimental design has a wide and diverse literature spread across
several disciplines, and some of these topics are being addressed in
current research under such titles as sequential design, active data
selection, and active, adaptive, or incremental learning.

\emph{Evolving goals for inference.}  In both examples the goal was
parameter estimation, the model being given.  In reality, the goals of
inference may not be so clear-cut.  In exoplanet surveys, observers will
often not be sure a system has a planetary companion at the start of an
exploration, so the goal will initially be detection of a planet.  Or if a
system is targeted because it is known to have a planet, the goals may
include detection of possible additional planets.  At some point, the goal
may shift from detection to estimation.  In landmine detection, the goals
will similarly shift from target detection to target classification (to
distinguish rocks,
debris, antipersonnel mines, and antitank mines), with parameter estimation
largely subsidiary to these goals.  How do design criteria for detection
compare to those for estimation?  When and how should the adaptive
methodology shift its goal from detection to estimation?  The work of
Toman \cite{Toman96}\ on Bayesian design for multiple hypothesis testing
provides a starting point for addressing these questions.

\emph{Generalizing the utility function.}
Our utility function was simply the information provided by new data.
In some settings, one may wish to incorporate other elements in the
utility function, such as the cost of observing as a function of time
or sample size.  How can a scientist map such costs to an information
scale so that information and other costs or benefits can be combined
into a single utility function?  

\emph{Computational algorithms.}
We used a simple rejection method for generating posterior samples in
our example.  While attractively simple, such an
approach is not useful for problems with more than five or six
parameters (even fairly sophisticated envelope functions will waste too
many samples).  The obvious tool for addressing this is MCMC, but the
Markov chain must ultimately sample over both the parameter space and
the sample space (of future observations).  Are there MCMC algorithms
uniquely suited to adaptive exploration?  M\"uller and Parmigiani and
their colleagues \cite{MP95,CMP95,MP96,Muller99}\ have developed a variety of
Monte Carlo approaches to Bayesian design in various settings that
should be helpful in this regard, though, as here, they have so far
treated fairly simple cases.  Also, since adaptive exploration
offers the hope of quickly reducing uncertainties, at some point it may
make sense to linearize about the best-fit model and use analytic
methods.  Criteria need to be developed to identify when this is
useful.

\emph{Design for the ``setup'' cycle.}
In our examples, the observing strategy for the first cycle was chosen
somewhat arbitrarily.  Ideally, it would be chosen using design
principles and prior information.  This raises many practical and theoretical
questions.  What should the size of a ``setup'' sample be?  Should
adaptive exploration start after a single sample, or are there benefits
(perhaps associated with computational complexity) for starting with larger
samples?  Can the algorithms used for analysis when several samples
are available also be used for designing the setup strategy, or are
different algorithms required if prior information is very vague?
Clearly, there is overlap between these issues and those already raised.
This kind of design issue has been addressed informally for planning
observations for the Hubble Space Telescope Cepheid key project
\cite{Cepheid94}.  Can a more formal approach improve on such a priori designs?

\emph{When is it worth implementing?}
Finally, though BAE provided impressive gains in the examples, such
dramatic levels of improvement cannot be expected generally.  Criteria
must be identified to help determine when BAE or other formal design
approaches may be useful.  Measures of model nonlinearity and posterior
nongaussianity may prove useful here.  Also, it seems likely that the
success of the approach depends critically on accurate model
specification.  Studies of robustness are needed to ascertain how
performance might degrade with model inaccuracies, and whether the method
can mislead investigators by avoiding parts of the sample space where one
should search for departures from the model predictions (e.g., regions of
the sample space where a model's predictions vary only very weakly with the
parameters).  Finally, is the approach ever valuable in settings with
highly flexible models (e.g., semiparametric and nonparametric models)?

We hope this brief introduction will encourage scientists, engineers and
statisticians to explore these issues together in a variety of
contexts.


\begin{theacknowledgments}
I thank David Chernoff for spurring my interest in adaptive design
for \SIM, for valuable discussions of the ideas presented here, and for coming up
with the term, ``Bayesian adaptive exploration.''  I also thank Paul 
Goggins for suggesting the landmine example and for valuable discussions
about this example and other aspects of BAE.
This work was supported by NASA through
grant NAG5-12082 and through the SIM EPIcS Key Project.
\end{theacknowledgments}


\bibliographystyle{aipproc}   

\bibliography{biblio}

\IfFileExists{\jobname.bbl}{}
 {\typeout{}
  \typeout{******************************************}
  \typeout{** Please run "bibtex \jobname" to optain}
  \typeout{** the bibliography and then re-run LaTeX}
  \typeout{** twice to fix the references!}
  \typeout{******************************************}
  \typeout{}
 }

\end{document}